\documentclass{sigchi}

\CopyrightYear{2016}
\setcopyright{acmlicensed}
\doi{http://dx.doi.org/10.475/123_4}
\isbn{123-4567-24-567/08/06}
\conferenceinfo{CHI'16,}{May 07--12, 2016, San Jose, CA, USA}
\acmPrice{\$15.00}

 \toappear{
	To appear at UIST 2020.
 }

\usepackage{balance}       
\usepackage{graphics}      
\usepackage[T1]{fontenc}   
\usepackage{txfonts}
\usepackage{mathptmx}
\usepackage[pdflang={en-US},pdftex]{hyperref}
\usepackage{color}
\usepackage{booktabs}
\usepackage{textcomp}
\usepackage{xac}
\usepackage{float}
\usepackage{url}

\usepackage{microtype}        
\usepackage{ccicons}          

\usepackage{todonotes}

\def\plaintitle{Geno: A Developer Tool for Authoring Multimodal Interaction on Existing Web Applications}

\def\emptyauthor{}
\def\plainkeywords{Multimodal interaction; voice input; developer tool; natural language processing}

\makeatletter
\def\url@leostyle{%
  \@ifundefined{selectfont}{
    \def\UrlFont{\sf}
  }{
    \def\UrlFont{\small\bf\ttfamily}
  }}
\makeatother
\def\pprw{8.5in}
\def\pprh{11in}

\definecolor{linkColor}{RGB}{6,125,233}
\hypersetup{%
  pdftitle={\plaintitle},
  pdfauthor={\emptyauthor},
  pdfkeywords={\plainkeywords},
  pdfdisplaydoctitle=true, 
  bookmarksnumbered,
  pdfstartview={FitH},
  colorlinks,
  citecolor=black,
  filecolor=black,
  linkcolor=black,
  urlcolor=linkColor,
  breaklinks=true,
  hypertexnames=false
}

\begin{document}

\title{\plaintitle}

\numberofauthors{3}
\author{%
 \alignauthor{Ritam Jyoti Sarmah\\
    \affaddr{UCLA HCI Research}\\
    \email{rsarmah@g.ucla.edu}}\\
 \alignauthor{Yunpeng Ding\\
    \affaddr{UCLA HCI Research}\\
    \email{dyp1225@g.ucla.edu}}\\
\alignauthor{Di Wang\\
    \affaddr{UCSD Computer Science}\\
    \email{diwang0503@gmail.com}}\\    
 \alignauthor{Cheuk Yin Phipson Lee\\
    \affaddr{UCLA HCI Research}\\
    \email{phipsonleecy@gmail.com}}\\
\alignauthor{Toby Jia-Jun Li\\
    \affaddr{CMU HCII}\\
    \email{tobyli@cs.cmu.edu}}\\
\alignauthor{Xiang `Anthony' Chen\\
    \affaddr{UCLA HCI Research}\\
    \email{xac@ucla.edu}}\\
}

\maketitle

\begin{abstract}

Supporting voice commands in applications presents significant benefits to users. However, adding such support to existing GUI-based web apps is effort-consuming with a high learning barrier, as shown in our formative study, due to the lack of unified support for creating multimodal interfaces. We present Geno---a developer tool for adding the voice input modality to existing web apps without requiring significant NLP expertise. Geno provides a high-level workflow for developers to specify functionalities to be supported by voice (intents), create language models for detecting intents and the relevant information (parameters) from user utterances, and fulfill the intents by either programmatically invoking the corresponding functions or replaying GUI actions on the web app. Geno further supports multimodal references to GUI context in voice commands (\eg ``move this [event] to next week'' while pointing at an event with the cursor). In a study, developers with little NLP expertise were able to add multimodal voice command support for two existing web apps using Geno.
\end{abstract}




\section{Introduction}


The advent of data-driven speech recognition and natural language processing (NLP) technology holds the promise of enabling robust and intelligent voice input that can recognize users' intent from natural language expression.

Meanwhile, as more applications become ubiquitously available on the web, adding multimodal, voice-enabled input on existing web apps presents important benefits. Voice input enhances GUI web apps' accessibility for visually-impaired users. Voice+GUI multimodal interaction also adds to the expressiveness of singular input modality \cite{cohen1997quickset,laput2013pixeltone,gao2015datatone,setlur2016eviza, DBLP:conf/sui/KangGL0C19}.


However, currently it takes a significant amount of work to augment existing web apps to support voice+GUI input. Despite the availability of multiple existing APIs and toolkits (\eg Chrome, Mozilla, W3C, Annyang, Artyom), our formative studies with five web developers identified the following barriers:
\one the amount of new code to write, including the effort of refactoring the existing codebase;
\two the gap of NLP expertise---to realize the NLP capability of a voice input, non-expert developers often find it laborious and challenging to develop mechansisms for understanding natural language inputs.
\three the lack of unified, integrated support for creating multimodal interaction---developers find it hard to `map' the development of voice input to their familiar GUI building paradigm and are unfamiliar with the best practice.


To ease the addition of voice input to GUI, prior research focused on enabling end-users to create custom voice assistants on personal devices, \eg smartphones \cite{li2017sugilite,li2018kite,li2019pumice}. While the end-user approach benefits users without significant technical expertise, it requires substantial efforts from \textit{each} end user of the app, which does not easily scale up.  
To complement the end-user-oriented approach, we focus on developer tools that can help developers make voice input readily available on existing web apps for all end-users to ``walk up and use''.
Further, beyond prior work that considered voice input as a separate modality, we want to support developers to integrate existing inputs (\eg mouse) multimodally with voice.

To achieve these goals, we develop Geno---a developer tool for authoring voice input single- or multimodally on existing web apps. 
Different from end-user tools that often revolve around demonstration at the front-end, Geno assumes that a developer is familiar with their own codebase at the back-end.

\subsection{Scenario Walkthrough}
\fgref{overview} shows an overview of the Geno IDE that consists of a file explorer, a code editor, and a preview of the web app. 

\fgw{overview}{overview}{1}{
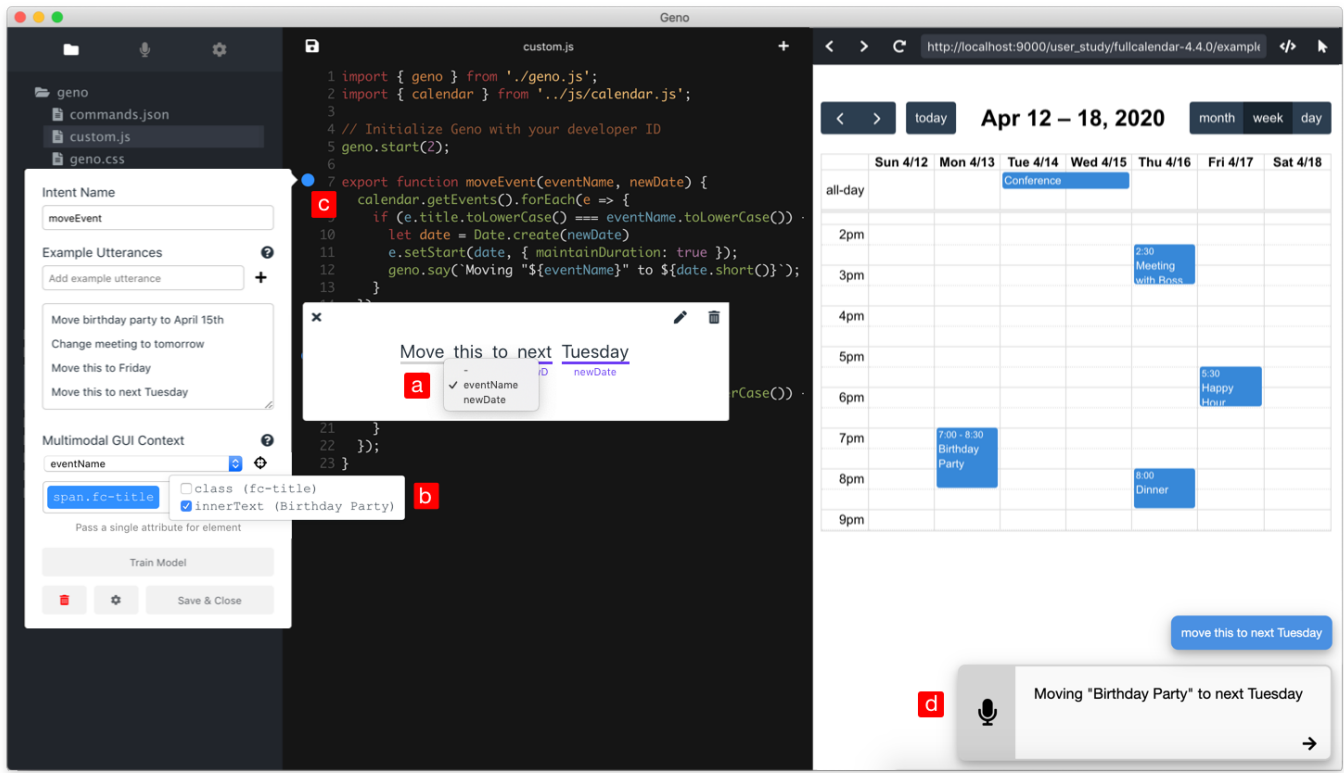
    In a few steps, Geno enables a developer to create a multimodal voice+GUI input where a user can point at an calendar event and say when to move the event (d): \#1 Specifying Target Action (\eg selecting the \texttt{moveEvent} function as the intent) (c);
    \#2 Configuring Voice Input by providing and labeling example utterances (\eg ``this'' is the value for the \texttt{eventName} parameter) (a)
    \#3 Adding GUI Context by demonstration (\eg hovering an event element where its \texttt{innerText} attribute will also be used for \texttt{eventName}) (b).
}

We demonstrate an exemplar workflow using Geno 
for a calendar app\footnote{\url{https://fullcalendar.io/}}, where a specific `view' often limits the direct manipulation of events. 
For instance, in the `week view' (\fgref{overview}), it is cumbersome to drag a calendar event to a different week or month.
A developer wants to create a multimodal voice+GUI input to ease this interaction, \eg one can hover the mouse over an event and say ``Move this to next Tuesday''.

To achieve this task, Geno supports the developer to demonstrate possible multimodal inputs, identify relevant information from those inputs (called parameters), and use such parameters to programmatically invoke specific actions.

First, the developer provides an intent name and some sample utterances, and labels the parameter values in the utterances (\fgref{overview}a). 
For example, as shown in \fgref{overview}c, for the intent ``moveEvent'' the developer provides an utterance ``Move this to next Tuesday'', and labels the entities (extracted by Geno)---`this' as \texttt{eventName}, and `next Tuesday' as \texttt{newDate}.

Next, the developer adds `GUI context'---a non-voice modality---as an alternate means of specifying \texttt{eventName}
by clicking any event element 
As shown in \fgref{overview}c, Geno automatically identifies the clicked element and displays its attributes as a checklist. The developer checks the \texttt{innerText} property to be used as the value for \texttt{eventName}.

Finally, the developer implements a simple wrapper function called \texttt{moveEvent}, whose skeleton is automatically generated in the current JavaScript file (\fgref{overview}c) to make use of the two multimodal parameters extracted by Geno. 
Note that this step is specific to the internal logic of the developer's web app; by design, Geno maintains a loose coupling with such logic and does not attempt to automate this step.

At run-time, Geno floats over the calendar app (\fgref{overview}d). Clicking the \faMicrophone \ icon (or using a keyboard shortcut) starts a multimodal voice+GUI input. Geno automatically recognizes the intent from the utterance, which leads to the target action of \texttt{moveEvent}. Geno will look for the two requisite parameters: \texttt{newDate} from the utterance and \texttt{eventName} from mouse events. 
If a parameter cannot be found, Geno will automatically prompt the user to manually specify the information.

%





\subsection{Validation \& Contributions}
We conducted an hour-long developer study with eight participants, who were able to learn and use Geno to implement multimodal voice+GUI input on existing calendar and/or music player apps. 
Following Geno's high-level workflow, participants found it much easier to build voice/NLP-enabled interaction than they had previously expected.
Meanwhile, it remained at times challenging to make certain design decisions (\eg to use function or GUI demonstration as the target action, which element/attribute to use as GUI context), which could be better supported in the future with a comprehensive tutorial and support for {\it in situ} testing and visualization.


Overall, Geno makes a tool contribution to the development of multimodal input. Specifically,

\begin{itemize}
    \item Geno embodies existing techniques for processing multimodal input while abstracting low-level details away from developers who might not be an expert;
    \item \rev{Geno unifies the creation of different modalities into a single workflow, thus lowering the complexity of programming an interaction multimodally};
    \item Geno works on existing web apps, whose accessibility can be enhanced by adding complementary input modalities;
\end{itemize}



\section{Related Work}
We first review prior work on multimodal interaction techniques and applications and then discuss existing tool support for authoring multimodal interaction.

\subsection{Multimodal Interaction Techniques \& Applications}
Multimodal interaction leverages the synergic combination of different modalities to optimize how users can accomplish interactive tasks \cite{oviatt1999ten}. One seminal work was `Put-that-there', whereby a user can simply point at a virtual object and speak to the system to place that object by pointing at a location on a 2D projected display \cite{bolt1980put}. QuickSet demonstrated multimodal input on a military planning interface that allows a user to pen down on a map and utter the name of a unit (\eg ``red T72 platoon'') in order to place it at that location \cite{cohen1997quickset}. 


Prior work on the design of multimodal \cite{tse2007multimodal, oviatt2000designing} interaction motivates our second goal of combining an additional voice modality with existing GUI-based input: users can offload a subset of GUI-based tasks that can be more optimally accomplished via voice.
To understand such complementary relationships between modalities. Oviatt found that users' natural language input is simplified during multimodal interaction, as the complexity is offloaded to more expressive modalities \cite{oviatt1999mutual}. Cohen \etal pointed out that direct manipulation can effectively resolve anaphoric ambiguity and enable deixis commonly found in natural language communication \cite{cohen89synergic}. 

Voice-enhanced, multimodal interaction has been explored in a plethora of application domains. 
Xspeak uses speech to manage windows in the XWindow System, such as saying the window names to switch between windows in focus
\cite{schmandt1990augmenting}.
Voice input can also ease tasks with a large number of GUI menu options such as image editing by directly speaking out the command \cite{laput2013pixeltone}.
On the other hand, when the content of the interaction is heavily loaded with information (\eg visualization of a complex data set), multimodal interaction allows the use of natural language to describe a query and direct manipulation to resolve ambiguities \cite{gao2015datatone,setlur2016eviza}.
The recent focus on automotive UI also suggests a promising scenario for introducing multimodal interaction. For example, SpeeT is a speech-enabled, multi-touch steering wheel that allows drivers to use speech to select a in-car functionality and gestures to adjust the settings \cite{pfleging2011speet}.
Interacting with an Internet of Things can also benefit from a combination of freehand pointing and voice commands, as demonstrated in the Minuet system \cite{DBLP:conf/sui/KangGL0C19}.
\rev{As Web-based UI becomes increasingly pervasive, it becomes important to offer flexible moddalities at the users' disposal, \eg using input targeting \cite{swearngin2017genie};
we expect our tool to be extensible to address multimodal interaction in these above application scenarios.}

\subsection{Tool Support for Authoring Multimodal Interaction}
Previous research on tools for authoring multimodal interaction spans three approaches: architectural support, specification language, and end-user development.

{\bf Architectural support.}
The Open Interface Framework provides both a kernel and a development environment to allow for heterogenous, distributive interactive components of different modalities to compatibly run as an integral system \cite{serrano2008openinterface, mcgee2009user}.
Bourguet separated the development of multimodal interaction into \one specifying the interaction model using a finite state machine; and \two using a multimodal recognition engine to automatically detect events relevant to the interaction model \cite{bourguet2002toolkit, bourguet2003designing}.
HephaisTK is a toolkit where recognition clients observe input from different modalities, which is integrated by fusion agents and can be subscribed by client applications \cite{dumas2009hephaistk}.
mTalk is a multimodal browser that provides integrated support (\eg cloud-based speech recognition) to ease the development of multimodal mobile application \cite{johnston2011mtalk}. Speechify is a toolkit that wraps speech recognition, speech synthesis, and voice activity detection APIs to assist with the rapid construction of speech-enabled Android apps~\cite{kasturi2015cohort}.
Little of this work, however, supports the development of user interactions at the front end, which our research primarily addresses in Geno.

{\bf Specification language}
has been explored extensively in prior work.
Obrenovic and Dusan described an UML-like language to represent multimodal interaction that can be generalized or transferred between different designs and analyses \cite{obrenovic2004modeling}.
XISL is one of the earliest XML-based markup languages that focused on describing the synchronization of multimodal inputs/outputs, as well as dialog flow and transition \cite{katsurada2003xisl}.
MIML presents an important feature that creates three layers of abstraction when specifying multimodal interaction: task, interaction, and device \cite{araki2009multimodal}.
UnisXML \cite{stanciulescu2005transformational} and TERASA \cite{paterno2008authoring} take a similar transformational approach, \ie deriving an instance of multimodal interaction from abstractions of task, domain and user interfaces.
Emma focused on capturing and annotating the various stages of processing of users’ multimodal inputs \cite{johnston2009emma}.
Different from all this work, Geno provides an interactive authoring tool that overcomes the lack of directness and expressivity of using a specification language.

{\bf End-user development} 
of multimodal interaction has become a promising option for end users to develop the support for their own desired tasks. Multimodality is often used to provide ``naturalness'' in the development process~\cite{myers_making_2017} to make it closer to the way the users think about the tasks~\cite{green_usability_1996}.
Sugilite allows end-users to create voice-activated task automation by demonstrating the task via directly manipulating existing app GUIs \cite{li2017sugilite}. 
Unlike Geno, Sugilite lets end users enable their personal tasks to be invoked by voice commands. Although the ``programming'' process is multimodal, the invoking of automation uses only voice commands. In comparison, Geno is a developer tool that helps developers to quickly add the support for mutlimodal interactions for existing web GUIs. 
Pumice builds off of Sugilite and adds support for learning concepts in addition to procedures \cite{li2019pumice}.
Improv allows end-users to replace direct manipulation with indirect gesturing on a second device, employing a programming by demonstration (PBD) approach to record and programmatically replay input events across devices to a web app \cite{chen2017improv}.
Appinite uses a multimodal mixed-initiative mutual-disambiguation approach to help end-users clarify their intents for ambiguous demonstrated actions, addressing the challenge of ``data description'' in PBD \cite{li2018appinite}.

\rev{Finally, there are commercial tool support, \eg Amazon Lex, which mainly focuses on the speech modality manifested as chat bots, rather than enabling multimodal interaction.}

\section{Formative Study}
Following the approach for designing developer tools~\cite{myers2016programmers}, we conducted a formative study to identify challenges in developing multimodal interaction and the gap of existing toolkits/libraries/APIs in their support of such development.

\subsection{Participants}
We recruited five participants (four male, one female, aged 19-23) from a local university majoring in Electrical Engineering and Computer Science (four undergraduate and one graduate students). All participants considered web development (JavaScript/HTML/CSS) as their skills, although their experiences varied, ranging from six months to two years. Only one participant reported having tried an API related to speech recognition and NLP, while others reported no related experience at all. Each participant received a \$20 Amazon gift card for compensation.

\subsection{Procedure \& Tasks}
Participants were randomly assigned to one of these two tasks of adding voice command capabilities to existing web applications based on existing open-sourced codebases:

\begin{itemize} [leftmargin=0.25in]
    \item Task option 1--Calendar: create a voice command to add an event at a given date and time\footnote{\url{https://github.com/fullcalendar/fullcalendar}};
    \item Task option 2--Music player: create a voice command to play/pause music\footnote{\url{https://github.com/521dimensions/amplitudejs}};
\end{itemize}

Participants were allowed to use any IDE and third-party toolkits/libraries/APIs; in case they could not find any, we also provided a list of five commonly used voice input related resources:
W3C\footnote{\url{https://w3c.github.io/speech-api/}}, Mozilla\footnote{\url{https://developer.mozilla.org/en-US/docs/Web/API/SpeechRecognition}}, Chrome\footnote{\url{https://developers.google.com/web/updates/2013/01/Voice-Driven-Web-Apps-Introduction-to-the-Web-Speech-API}}, Artyom\footnote{\url{https://sdkcarlos.github.io/sites/artyom.html}} and Annyang\footnote{\url{https://www.talater.com/annyang/}}.

After a brief introduction, each participant was given 60 minutes to perform the development task, during which the experimenter asked them to think aloud. In the end, participants were asked to complete a short questionnaire to evaluate their overall experience and performance in the specified task.

\subsection{Findings \& Design Implications}
All participants were unable to complete the assigned task of adding the support for voice input to a provided web app within 60 minutes. Admittedly, such an outcome could have been a result of two obstacles: a lack of web programming expertise in general and the unfamiliarity to voice input and NLP.
Below we discuss our observations specific to the latter obstacle and the corresponding design implications for Geno.

\one Due to a lack of NLP expertise, all participants struggled to come up with a programmatic way of parsing voice inputs. Most attempted to manually create `hard-coded' rules that only worked for specific cases.
Some APIs used by the participants (\eg W3C Speech Recognition API) only perform speech recognition but no parsing; some others (\eg Annyang) do provide parsing capabilities but requires knowledge of using a regular expression.\\ 
\insec{Design implication:} the parsing of the transcribed voice input should be automatic and require as little extra knowledge and effort as possible.


\two Participants found the process of programming voice input unnatural and unfamiliar compared to their experience of developing non-voice GUI. Specifically, P5 suggests ``defining a single uniform function'' that could be reused across different modes of input. P5 explained that, by doing so, he could use his knowledge of, \eg button event listeners for voice input.\\ 
\insec{Design implication:} instead of considering voice as a different mechanism, Geno should provide a unified process for developers to realize an interactive task across any combination of voice and pointing modalities.

\three None of the participants attempted or were able to develop an error handling mechanism---that is, when an intended voice input is not recognized by an API, what then? Foremost, participants were often confused about why certain utterance was not recognized as the intended interactive behavior. By default there is no feedback from the API whether the app is listening at all, or if so, what is heard and how does that (not) trigger the intended action of the application.\\ 
\insec{Design implication:} Three levels of feedback are useful for developing NLP-enabled interaction: whether the app is listening, what is heard and what is (not) being acted upon; further, there should be support for developers to handle misrecognition errors, \eg using a dialog to ask for disambiguation or extra information from the user when ambiguity occurs. 


\four Last but not least, we were surprised to find that three participants at the beginning were bogged down by the inability to access the microphone on the experiment laptop. P3, P4, and P5 experienced issues with the microphone and audio interface. P4 struggled with permission issues, and P5 had runtime errors due to an undetected microphone, which he later discovered was caused by incorrect declaration and misplacement of the voice recognition object.\\ 
\insec{Design implication:} although not the core of multimodal input, low-level hardware access should be integrally supported as part of the Geno toolkit and should be abstracted away from developers' main workflow.

\section{Geno: Adding Multimodal Input to Existing Apps}
In this section, we describe how Geno works: 
\one how to create an intent at development time and 
\two what is the run-time behavior of an intent. 
Specific implementation details are discussed in the next section.

\subsection{Development Time}
With Geno, the main workflow of adding an intent to an existing web app is guided by a dialog (\fgref{dialog}) that involves the following steps: 
specifying target action, configuring voice input, and specifying GUI context.




\subsubsection{Step \#1: Specifying a Target Action}
\label{step1}
Geno supports adding a voice modality to two types of target actions—an intent can invoke either a JavaScript function, or a sequence of GUI interactions on the interface of the web app. 
Sometimes these two types are equivalent (\eg calling a \texttt{changeColor} function \vs using a color picker widget), but in general, GUI interactions allow for one-off, nonparametric input (\ie simply using voice to invoke the same input sequence, such as clicking on the ``next song'' button) whereas functions' behavior can be varied based on specific contents in the voice input (\eg moving a calendar event to ``next week'', ``Friday'' or ``tomorrow'').

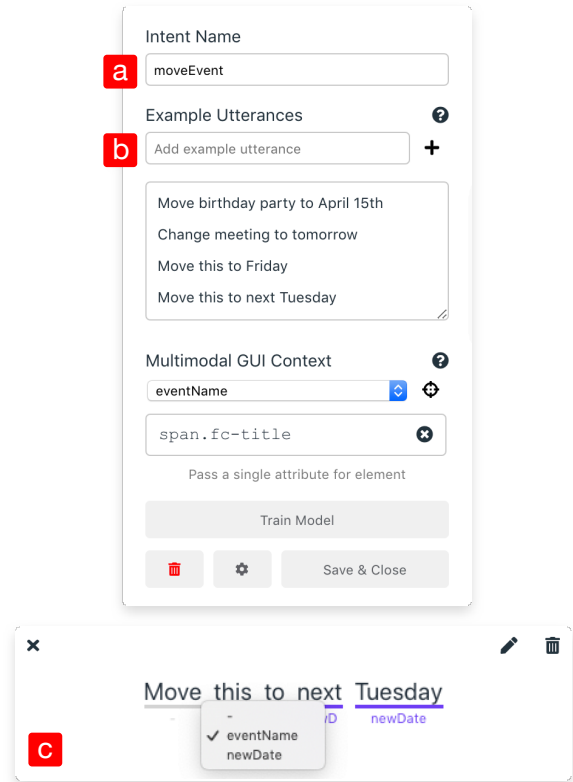
\fg{dialog}{dialog_alt}{0.9}{An overview of Geno's workflow, which is guided by this dialog: after specifying a target action, a developer creates an intent (a), provides example utterances (b), labels extracted entities as parameter values (c), and optionally adds GUI context to supplement the voice modality. 
\vspace{-1em}
}

{\bf To specify a target action that executes an existing function}, a developer clicks the \faCircle~ next to a function (\fgref{create_fn}a), which creates a new intent and opens a dialog for configuring the voice input with the function/arguments pre-filled as the target action/slots (details in Step \#2 below). 

It is possible that a single function for an intended target action does not yet exist. For example, as shown in \fgref{overview}, moving an event to a new date in the calendar app requires invoking multiple existing functions to retrieve an \texttt{Event} object from the title of the event, create a \texttt{Date} object from the user's natural language description of the target date, and set the \texttt{date} field of the \texttt{Event} object using the \texttt{setDate()} function\footnote{Note that programming such logic is application-specific, thus not automated by Geno.}.  In this case, Geno allows a developer to create 
an intent without a specific function by clicking the \faPlus \ button 
(\fgref{create_fn}b) to initiate the same dialog, go through the following steps (\#2 and \#3), and then
Geno automatically generates a skeleton of the new function with the function name and its list of arguments so that the developer can implement the body of the function. In a function, the developer can also use Geno to say the result of the action by speech using the \texttt{geno.say()} function.

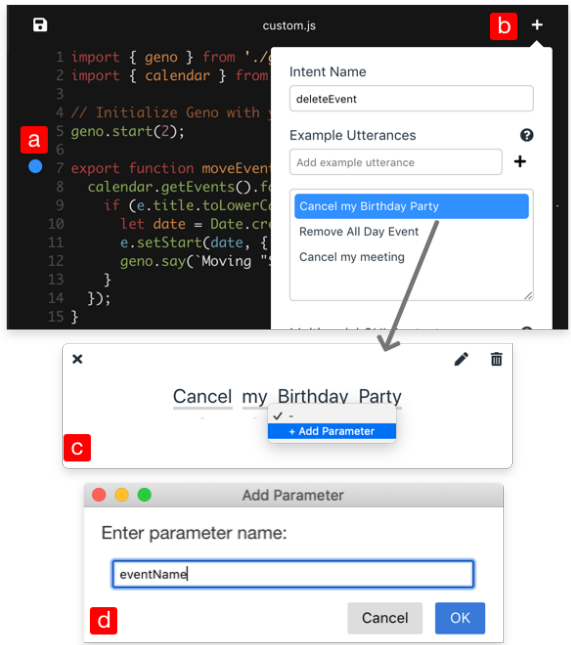
\fg{create_fn}{create_fn}{0.9}{\#1 Specifying a target action as a function, which can be one that already exists (clicking the \faCircle~ aside a function) (a), or a new one (clicking the \faPlus~ button) (b). For a new function, Geno allows a developer to create a new parameter, using entities in an utterance as the arguments of the new function.
\vspace{-1em}
}

Currently, Geno only supports associating voice input with one function at a time; in the future, we plan to explore multiple functions and their different relationships with respect to a voice input (detailed in discussion).



\fgw{create_demo}{create_demo}{1.0}{
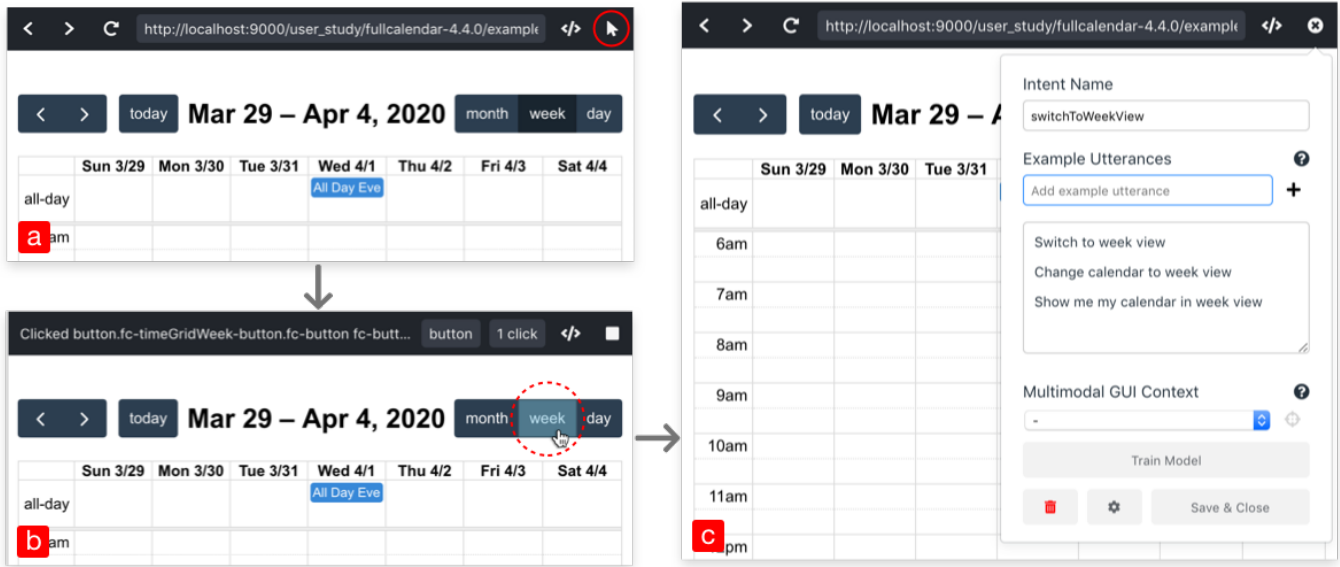
  \#1 Specifying a target action (by demonstrating GUI interactions): clicking the \faMousePointer~ button starts the recording of a sequence of input events and recognizes clicking the `week' button as a click action (b), based on which an intent is created to trigger the replay of the recorded click action (c).
}

{\bf To specify a sequence of GUI interactions as a target action}, a developer would instead focus on the web app preview and click on the \faMousePointer~(\fgref{create_demo}a) button to demonstrate a sequence of actions on the GUI, such as clicking the `week' radio button to switch the calendar's view (\fgref{create_demo}b), or to open a color picker and select the color red. 
After the developer enters the demonstration mode, Geno displays a blue highlight overlay above the element currently hovered over by the mouse cursor and a description of this element 
to help the developer identify the correct target element (\fgref{create_demo}b). 

\subsubsection{Step \#2: Configuring Voice Input}


The next step in creating an intent to configure its voice input. This includes
\one providing sample natural language utterances for invoking this intent; and
\two labeling the task parameter values in the sample utterances.

To add new sample utterances (\fgref{dialog}b), a developer types in \rev{two to three} example utterances, \ie different ways of saying a command to invoke this intent, such as ``reschedule this to next week'', ``move {\it Birthday Party} to 6PM today'', ``shift {\it Group Meeting} to Friday''. 
Here we chose typing in rather than speaking out sample utterances, as typing makes editing easier while composing an utterance. 
These sample utterances are used for training an intent classification model that, at run-time, can associate the user's voice commands with the corresponding intents (details in the Implementation section).

After adding the sample utterances, the developer labels the parameter values in those utterances. As shown in \fgref{dialog}c, for each utterance, the developer can click on a part of the utterance (\eg ``Tuesday'') and label it as the value for a parameter (e.g., ``\texttt{newDate}''). 
If the intent was created from an existing JavaScript function, then the list of parameters will be automatically populated with the list of arguments in the function (\fgref{overview}a). 
Otherwise, if the target function has yet to be implemented, the developer needs to define new parameters as they label the parameter values (\fgref{create_fn}cd). 
Those labels are used for training an entity extraction model for extracting task parameter values from the user's run-time voice commands (details in the Implementation section). Note that a target action can be nonparametric (\eg when the target JavaScript function does not have any arguments or when the target action is a sequence of GUI interactions), in which case the developer can skip labeling parameter values.

\fg{create_context}{create_context}{0.9}{
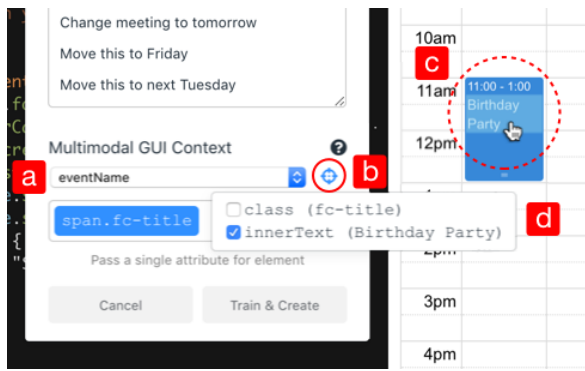
  \#3 Adding GUI context: first select a parameter \texttt{eventName} (a), then start a demonstration (b) and click or drag to select a calendar event element (c) and its attribute \texttt{innerText} which can be used as a value for \texttt{eventName}.
  }

\subsubsection{Step \#3: Adding GUI Context}
Geno supports multimodal inputs that use both voice inputs in natural language and mouse inputs of referencing to GUI context. For example, a user might say ``reschedule this to next week'' while pointing to an event on the calendar with the cursor. In this case, the reference of ``this'' specifies the value of the parameter \texttt{eventName} in this intent (\texttt{moveEvent}). Currently, Geno supports extracting such information from GUI contexts either as a single HTML element hovered over by the cursor or multiple elements marquee-selected by dragging. Multi-selected elements are extracted as lists, which is useful for supporting commands such as ``adding \textit{all of these} to the playlist'' while selecting multiple songs.

To add the support for GUI context input, the developer first selects a target parameter (e.g., \texttt{eventName}) from the ``Multimodal GUI Context'' dropdown menu (\fgref{create_context}a), and then clicks on the \faCrosshairs~ icon (\fgref{create_context}b) to demonstrate an example GUI context. As shown in \fgref{create_context}c,
in the calendar example, the developer first hovers the cursor over a calendar event, which highlights the HTML element as a potential selection, and clicking\footnote{Native clicking events on the web app were temporarily disabled while specifying GUI context.} an element would confirm it as GUI context.
Once an element is selected, Geno extracts a list of its HTML attributes such as its class name and innerText
(\fgref{create_context}d).
The developer then selects one target attribute that contains the desired value. In the calendar example, the developer chooses the \texttt{innerText} field, since it contains the name of the event. This demonstration allows Geno to create a data description query~\cite{li2018appinite} for the GUI context input, so that at run-time, when the user points to a GUI element or selects a list of GUI elements, Geno can extract the correct attributes from these elements and use them as parameter values in the detected task intent.

\subsection{Run-time}
At run-time, Geno floats over the application (\fgref{overview}d) on the user's end. A user can click the \faMicrophone~ icon or use a built-in keyboard shortcut (we use \texttt{Ctrl + \textasciigrave}) to start speaking a voice command, at times in tandem with references to GUI context.
Geno transcribes the user's speech to text and matches the text to an existing intent created by the developer. If Geno cannot find a matching intent, it will say ``Sorry, I didn't understand. Could you try again?'' to prompt the user to provide a new utterance.

In order to execute the target action of the matching input (\eg \texttt{moveEvent}), Geno will need to find the requisite parameters (\eg \texttt{eventName} and \texttt{newDate}). 
For each parameter:

\one If an entity corresponding to the parameter is found in the utterance, Geno will ``fill'' the target action with the entity;

\two If no entity is found, Geno first checks whether the GUI context has been configured for that parameter, and if the user has hovered over a matching GUI context with the voice command. If so, Geno extracts the parameter values from the user-specified GUI context (\eg the \texttt{innerText} of an HTML element for an event). 

\three If no GUI context is found, Geno will ask a follow-up question for each missing parameter, which the user can directly respond to,
\eg ``What is eventName?'', although the developer can customize the prompt question for each parameter in the options view in the popover (accessed by clicking \faCog~ button) where each parameter is listed with a text field to specify a custom question.


Once all the required parameters have been ``filled'', Geno executes the target action by either invoking the corresponding JavaScript function or replaying the sequence of demonstrated GUI interactions as configured by the developer.

\section{Implementation}
In this section, we describe key implementation details that underpin Geno's workflow of creating multimodal input on existing web apps. \fgref{diagram} shows an overview of Geno's system architecture.

\fg{diagram}{diagram}{1.0}{An overview of Geno's system architecture.}

\subsection{Natural Language Processing Pipeline}
Geno uses a typical frame-based dialog system architecture for processing the user's natural language commands. After the user speaks an utterance, Geno transcribes the speech to text using the Web Speech API's\footnote{\url{https://developer.mozilla.org/en-US/docs/Web/API/Web_Speech_API/}} SpeechRecognition toolkit. A natural language understanding (NLU) module then classifies the utterance into one of the intents defined by the developer and extracts the parameter values from the utterance. 

The intent classifier uses the StarSpace model~\cite{wu2018starspace}, which is a state-of-art general-purpose neural embedding model for text classification.
\rev{Given a voice query, the classifier returns a ranking of potential intent matches based on confidence of the match. We look at the match with the highest confidence, and execute the command if it meets a 50\% minimum threshold of confidence.}
The parameter value extractor uses a CRF entity extraction model\footnote{\url{https://rasa.com/docs/rasa/nlu/entity-extraction/}} to recognize named entities (\eg date, location) in the user utterance. Both models run on a remote server and communicate with the client-side of Geno and Geno-enabled web apps through HTTP requests. The models are implemented using the open-sourced Rasa library\footnote{\url{https://rasa.com/}}.

The intent classification and the parameter value extraction models are trained using the example utterances and their labeled parameters provided by the developer (Step~\#2). 
Specifically, the back-end first creates tokens from input sentences, encodes input sentences into vector representations, and then creates bag-of-word representations for model training.


For intents with parameters that support GUI context input, Geno's parameter value extractor looks for common demonstrative pronouns (\eg this, that) in the user's utterance. 
If found, the system will look for relevant HTML elements, extract attributes predefined by the developer (Step \#3) and fill in the corresponding parameter with the value from the GUI context. For example, if the user says ``Move \textit{this} to Friday'' while pointing at the event with the title ``Birthday Party'' for the Geno intent shown in \fgref{dialog}, Geno would detect the reference of ``this'' in the utterance, finds a calendar event element based on the hovering mouse position and retrieves its value ``Birthday Party'' from the event element's \texttt{innerText}, and uses it for the value of the task parameter \insec{eventName}. 

For each unfilled parameter in the intent, Geno asks a follow-up question until all the parameters for the intent are filled, so the intent is ready for execution. Geno uses Web Speech API's SpeechSynthesis library to ask these questions both by voice and by displaying the text in the slide-over window (\fgref{followup})


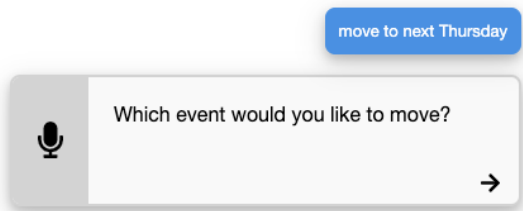
\fg{followup}{followup}{0.9}{If Geno cannot find a parameter value in any modality, it falls back to prompting the user to specify the parameter value verbally.
\vspace{-1em}
}


\subsection{Recording and Replaying GUI Demonstrations}

When the developer demonstrates GUI interactions as a target action (Step~\#1), Geno records mouse clicks and keyboard input. Although it is possible to click on any HTML element, not all of them represent meaningful actions intended by the developer. Thus Geno only records interaction with clickable elements by filtering out elements that do not have a \texttt{.click()} function. The recorded actions are saved as a list where each entry is a clicked element's tag and index, which can be used to retrieve the same element at run-time (\ie getting the \texttt{[index]}$^\text{th}$ element of type \texttt{[tag]} in the DOM tree) and programmatically invoke a \texttt{click} event on the element or enter text into the element.

\subsection{Extracting GUI Context}

The developer creates the support for using GUI context as a parameter by demonstrating in the web preview (Step~\#3). Geno identifies HTML elements as GUI context by detecting hovering over or dragging to marquee-select elements. Specifically, hovering is detected by when the mouse cursor displacements between consecutive \texttt{mousemove} frames are smaller than a threshold 
and the hovered element is retrieved using \texttt{document.elementFromPoint(x,y)}; dragging is detected when a displacement between \texttt{mousedown} and \texttt{mouseup} is greater than a threshold 
and elements completely within the rectangular dragged area will all be considered as GUI context.

Once the element(s) are located, Geno extracts the identifiable features for each selected element by its \texttt{tag}, \texttt{class}, and \texttt{element.attributes}. 
The developer can select one or more such features to create an attribute filter for this GUI context parameter (Step~\#3, Figure~\ref{fg:create_context}). This attribute filter is used for extracting the parameter value from the user's selected GUI elements at run-time. 
For example, when the developer clicks on the ``Birthday Party'' element in the calendar app to demonstrate the GUI context input for the parameter \texttt{eventName}, as shown in Figure~\ref{fg:create_context},  Geno extracts \texttt{fc-title} (the \texttt{class} of this element) and ``Birthday Party'' (the \texttt{innerText} of this element). The developer chooses the  \texttt{innerText} attribute to indicate using its value for the \texttt{eventName} parameter. At run-time, when the user hovers over a HTML element while saying a command that triggers the \texttt{moveEvent} intent, Geno finds the hovered HTML element that matches the selector \texttt{span.fc-title}, extracts the value of its \texttt{innerText} attribute, and uses this value for the parameter \texttt{eventName} when invoking the target JavaScript function for the intent \texttt{moveEvent}.

\subsection{Integrating with Existing Web Apps}


When a developer loads an existing web app into Geno, the system creates a directory in the root of the project folder that includes a \texttt{geno.js} file that supports all the requisite functionalities in Geno-enabled apps and a JSON file that contains all the developer-specified intents. 



After the developer specifies the intents and trains the NLP models, Geno automatically builds the project, updates the JSON file and imports \texttt{geno.js} and other requisite files into the web app. 
As shown in \fgref{diagram}, when a user runs a Geno-enabled app in the browser, \texttt{geno.js} gets executed,
which adds the Geno voice command button to the interface, manages the intents stored in the JSON file, communicates with the back-end server to understand the user's voice commands, and facilitates the execution of the target actions for user intents. Geno uses dynamic imports in JavaScript to execute functions with associated voice commands. 


\subsection{Environment \& Software Toolkits}
Geno uses Electron\footnote{\url{https://www.electronjs.org}} as the core for its IDE. Electron allows desktop apps to be built in JS and also supports cross-platform compatibility for macOS, Windows and Linux. 
Geno was written in TypeScript and React\footnote{\url{https://reactjs.org}} and uses
Acorn\footnote{\url{https://github.com/acornjs/acorn}} for parsing AST of JavaScript files to identify functions,
various React components for UI elements (\eg CodeMirror\footnote{\url{https://codemirror.net}} for code editor, Treebeard\footnote{\url{https://github.com/storybookjs/react-treebeard}} for file explorer),
Lowdb\footnote{\url{https://github.com/typicode/lowdb}} for database related tasks, and
Chokidar\footnote{\url{https://github.com/paulmillr/chokidar}} for watching files on disk and keeping UI components up to date for any database-specific state changes.

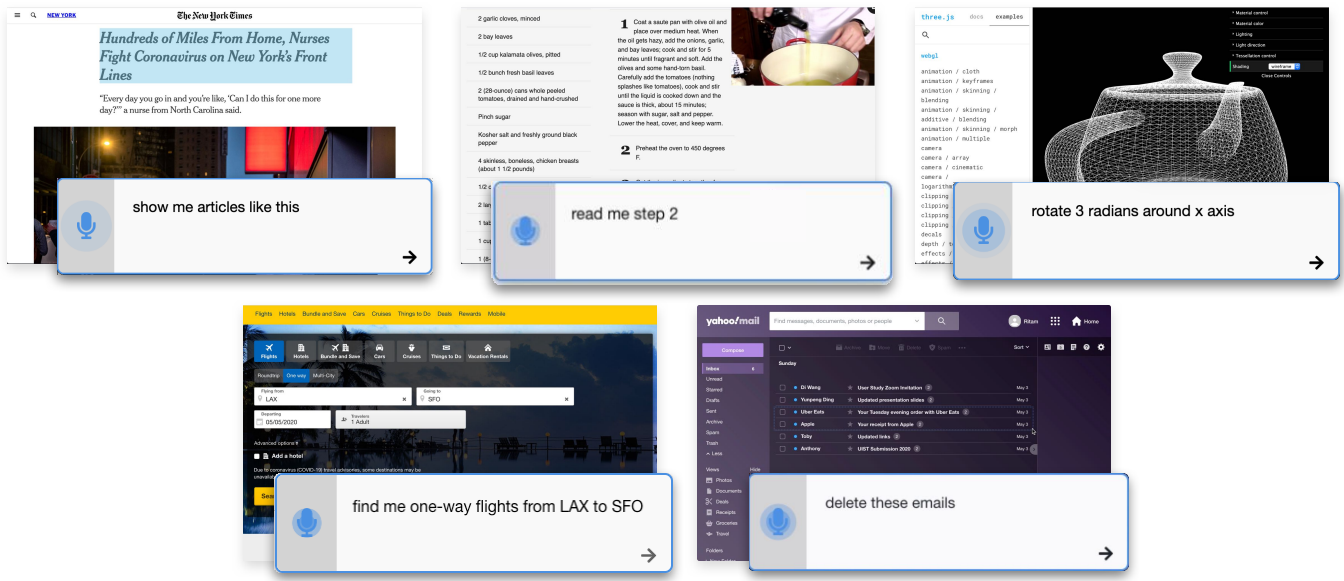
\fgw{examples}{examples}{1.0}{More voice+GUI input examples created using Geno based on actual websites.
\vspace{-1em}
}

\section{Existing Web Apps Multimodalized by Geno}
To demonstrate the expressiveness and practicality of Geno, we created multimodal input on five examples of existing actual websites and web apps, as shown in \fgref{examples}.
The New York Times: Search for related news articles on a website by hovering over interesting text;
FoodNetwork: Hands-free control of a recipe website while cooking, \eg play/pause or skip through a walkthrough instructional video and have Geno read out the steps;
Three.js: Switch between modes and manipulate 3D models by voice commands in 3D modelling software;
Expedia: Quickly search for flights by referring to airport codes;
Yahoo! Mail: Manage email more efficiently using multimodal input by dragging over emails and saying to ``Delete these'', or forwarding emails by hovering over an email and saying ``Forward this to Alex''.


\section{Developer Study}
We conducted a developer study \rev{with the research question to} validate the usability and usefulness of Geno for developers to create voice+GUI multimodal input on existing web apps.

\subsection{Participants}
We recruited eight participants aged 19-26 (mean 22.6, SD 2.1). All participants were male, five were students, and three were professional software engineers. We asked the participants to self-report their web development experience: four participants considered themselves expert in web development, one was intermediate and three were novice. Only one participant had prior experience developing voice interfaces. Each participant received a \$20 Amazon gift card for their time. 


\subsection{Apparatus}
We provided participants the same starting codebases as the ones used in the formative study.
Due to the COVID-19 pandemic, all study sessions were conducted remotely online via Zoom\footnote{\url{https://zoom.us}}. Specifically, we set up Geno in one experimenter's laptop and used Zoom to allow each participant to remotely control the laptop and interact with Geno to perform the development tasks.
We screen-recorded (including audio) the study sessions also using Zoom.

\subsection{Procedure \& Tasks}
Each one-hour study session consisted of a brief introduction to Geno, a walkthrough tutorial, two development tasks, a questionnaire, and a semi-structured interview.
We used a simple text editor app to introduce the goal of Geno and walked through an educational `Hello World' example where we showed each participant how to create voice input to change the color of text and to toggle the `bold' formatting button.

The main development tasks mirrored our formative study except that
for each task, we added an additional input that requires {\it both} voice and GUI inputs, and each participant was asked to complete tasks in {\it both} applications (as opposed to just one in the formative study).

\begin{itemize} [leftmargin=0.25in]
    \item Task 1--Calendar: create input to\\ 
    \one use voice to change to week view\\
    \two mouse-hover over an event and use a voice command to reschedule the event to $N$ days later;

    \item Task 2--Music player: create input to\\
    \one use voice to skip a track\\
    \two mouse-hover over a song and use a voice command to add the song to a playlist;
\end{itemize}

For each input in the assigned task, we provided the participant with five different test cases. For example, a test case for the second calendar input would be hovering the mouse over a ``Meeting'' event on Tuesday 10 AM this week and saying ``Postpone this by three days''.
After a participant finished developing each input, one experimenter tested the result by walking through and performing the pre-defined test cases.
We did not let the participants directly speak to Geno due to the often poor audio quality in the remote study, which affected the speech recognition performance. Instead, we asked the participants to test the app by telling the experimenter what to say, and the experimenter relayed the utterance for them.

At the end of the tasks, \rev{each participant} was asked to answer a questionnaire regarding the easiness and usefulness of using Geno. Based on their responses, we conducted a short semi-structured interview to understand where they felt the development was well-supported, what remained challenges and how they would see Geno to be improved in the future.

\subsection{Results \& Findings}
There were 2 multimodal interactions $\times$ 2 applications = 4 development tasks for each participant. Across our study, six users completed all four tasks in less than an hour, one user completed three tasks, and one completed two. The two developers who did not complete all the 4 tasks self-reported as novice web developers, and spent longer time writing and testing the JavaScript functions during the study. 


\rev{We conducted a thematic analysis of the qualitative data \cite{braun2006using}.}
Overall, Participants appreciated the mission of Geno: 
\inquo{I would definitely say it's very useful ... addressing the accessibility issue for most of the people who need it.}{7}

Participants also appreciated how Geno simplified the programming of multimodal input with a GUI-based guided workflow, achieving the \inquo{right balance of abstraction}{2}:

\quo{... provides like a GUI on how to adding interactions to your application.}{5}

\quo{It does so much for you. I just kind of forget that I actually have to do anything.}{2}

Participants found Geno overall easy-to-learn.
\quo{I just saw one example, and I was able to do it.}{8}
\quo{it was I think initially was a little hard, but I think I got it later ... the learning curve for this (Geno) is much better (than learning a library)}{6}

Importantly, Geno was considered \inquo{a very useful tool}{4} that
\inquo{definitely breaks down the barrier of traditionally you think adding anything machine learning related to your product is a very hard task.}{5}

{\bf NLU support is the most appreciated aspect of Geno}\\
Participants were overall impressed by how Geno made it much easier to program NLU into existing applications than what they had expected.

\quo{... adding a couple of examples and it knew exactly what you are talking about.}{1}

\quo{I definitely would not want to do it  myself so I would definitely be in favor of using something that would do most of it for me.}{2}

Specifically, despite a lack of NLP knowledge, participants were able to understand and learn how to provide example utterances and labeling entities:

\quo{... weren't necessarily like intuitive, but I mean, once you learned it, it was pretty usable.}{7}

\quo{... it was like a lot easier than I thought it was going to be. I thought I was gonna have to like do some complicated things, but I didn't really have to do anything that complicated.}{2}

Below we summarize the main challenges that the participants faced when using Geno to create multimodal input on existing web apps.

{\bf The challenge of specifying target function: function vs. GUI demonstration---when to use which}\\
\rev{We found participants sometimes confused about which approach to take.}
For example, when implementing multimodal input to move a calendar event, P1 first attempted to use demonstration, only to realize it was not possible to drag an event out of the current view. 
Participants also might not have realized that the support for demonstration currently is still limited to one-off action replay that cannot be parameterized.
We believe such confusion can be prevented with a tutorial with a set of comprehensive examples to explain the differences between specifying function and demonstrating GUI interactions, which we leave as future work.
We will also explore merging these two approaches, \eg automatically extracting function calls from GUI demonstration so that a developer can associate voice input to a specific function and enable parameterization.






{\bf The challenge of GUI context: which elements to choose (dev-time) and which elements to hover (run-time)}\\
Participants thought it was \inquo{convenient}{1} that Geno could automatically detect a UI element as the pre-defined GUI context. 
However, some participants had difficulty deciding which element to use as the GUI context. 
For the music player app, P4 chose the wrong element that did not contain information about the song (to be added to the current playlist); 
during testing, P4 forgot about which element s/he was supposed to hover. 
For not choosing the right element as the GUI context, we believe it is part of trial-and-error process typical in learning to use a new development tool, which can be better supported in the future, \eg iteratively testing multiple elements and their attributes {\it in situ} as a part of Step \#3.
For not knowing which element to hover at run-time, future versions of Geno can provide visual cues to indicate elements that have been pre-defined to serve as the GUI context.

{\bf The challenge of GUI context: scalability}\\
P7 pointed out that scalability could be another issue if the UI became more complicated, as hovering over an intended element would be challenging when there were multiple small surrounding or nesting elements.
To address the scalability issue, one possible solution for future work is to enable coarse selection of context: instead of having to hover over the exact element, Geno should allow for imprecise pointing and automatically search for the element that matches the type specified by the developer. 

{\bf The challenge of striking a balance between abstraction and transparency}\\
As Geno abstracts away details of multimodal input from non-expert users, it inevitably creates `black boxes' in the process where questions from developers might arise.
For example, when adding GUI context, P5 questioned the priority between processing voice and mouse inputs (we used a canonical frame-filling approach).
%
%
Even for the NLU part, multiple questions were raised. P1 wanted to find out the ``bound'' of the NLU model: \inquo{If I can understand the rule of NLU better, it can help...}{1};
P8 questioned how he could know when to stop adding utterances.
We believe in future work, Geno can provide more information to better inform non-expert users of Geno's process, \eg a dynamic visualization of the frame-filling approach, intermediating the performance of the NLU model as a developer provides more example utterances.
Importantly, lest to overwhelm developers with too much detail, Geno should allow developers to request such explanatory information on demand.
\section{Limitations, Discussion \& Future Work}
We discuss the current limitations of Geno that point to opportunities for future work.

\subsection{Specifying a Target Action}
When specifying a target action as a JavaScript function, Geno currently only supports using one function for each intent, although a developer can bypass this limitation by writing a wrapper function that invokes multiple subroutines.
In the future, we plan to explore the support for specifying multiple functions and their different relationships with respect to voice input.
For example, an action can consist of a sequence of function calls or calling different functions based on specific conditions.

When specifying a target action as a demonstration of GUI interactions, Geno currently only supports recording and replaying nonparametric clicking events.
In the future, we will support demonstrations that can be parameterized.
For example, a demonstration for ``Search for headphones'' would consist of typing ``headphones'' into a search box and clicking the search button.
In this case, the text entered can be considered a parameter that can be changed at run-time, \eg ``Search for speakers'' would replace the contents of the search box with ``speakers''.
Instead of simply replaying the recorded events, it is also possible to parameterize the sequence, which is demonstrated in Chen and Li's Improv framework \cite{chen2017improv} and will be explored in our future work.

Geno's mechanism of recording GUI interactions only supports clicking and text entry. It does not support interacting with widgets like sliders, knobs, date pickers, and spinners. It also does not record gesture inputs (drag \& drop, flicking etc.) The replaying mechanism relies on the tag and the order of HTML elements to locate the target element; therefore it may break for some dynamically-generated or adaptive web interfaces. In the future, we plan to extend Geno's support for various GUI widget types. We will also expand Geno's replaying mechanism to support using flexible queries to locate the target HTML element at run-time using approaches such as~\cite{li2018appinite}.






\subsection{Adding GUI Contexts}
Although Geno showcases the usefulness of GUI contexts, sometimes HTML elements' attributes cannot provide sufficient information for a specific parameter. For example, in the calendar app's case, the GUI context would not work if multiple events share the same title. Although Geno is able to read from the user's references to GUI context and use them for invoking target actions, it is up to the developer to decide whether and how it is possible to make use of such information. 

Another limitation is that Geno currently does not check the validity of parameter values when extracting them from voice utterances or GUI contexts. which may encounter exceptions at run-time when the user, for example, says ``move the event to [a non-date value]''. 
In future work, we plan to provide further exception handling and debugging support.
For example, exception handling code can be automatically generated. When an parameter value exception occurs, Geno will automatically fall back to manually ask the user for that parameter.

\rev{Further, it is also possible that a GUI element might be too small, making it difficult to specify as context. Future work can employ a increasingly broader range of searching to match users' GUI action to the specified contextual elements.}

\subsection{Supporting a Wider Range of Web Apps \& Platforms}
So far, we have only tested Geno on single-paged, desktop web apps \rev{, which is a reasonable starting platform for deploying Geno-generated multimodal interaction}.
Our future work will engineer the current implementation to support \rev{larger, more complex} multi-page web apps, \eg allowing for carrying over GUI context from one page to a function call that takes place on another.
To support mobile platforms, we will have to redesign how to obtain GUI context since there is no hovering (Input State 2) on touch screens. Dragging, on the other hand, is already supported on the mobile platform.
\rev{In addition, future work should reach out to more expert developers beyond our novice participants}

\rev{
\subsection{Model Generalizability}
Each NLU model is limited to a single application, since the model's intents are associated with functions unique to the codebase. However, in the future, we could create models for categories of applications (\eg calendar, food orders) that could be repurposed by multiple developers.
The GUI demonstration also requires additional information to generalize to different commands. For example, if a user has already defined a command for "move this meeting to next Tuesday" and would like to add a second question "move this meeting to next Tuesday and change duration to 1 hour", then a new rule would be required for the second question since it includes a new type of  information, \ie event duration.
}

\section{Acknowledgement}
This work was funded in part by the National Science Foundation under grant IIS-1850183. We thank our study participants and the reviewers for their valuable feedback.

\bibliographystyle{SIGCHI-Reference-Format}
\bibliography{main}

\end{document}